# Multiple-relaxation-time lattice Boltzmann modeling of incompressible flows in porous media

Qing Liu, Ya-Ling He*

*Key Laboratory of Thermo-Fluid Science and Engineering of MOE, School of Energy and Power Engineering, Xi'an Jiaotong University, Xi'an, Shaanxi, 710049, P.R.China*

(*Corresponding author. Email: yalinghe@mail.xjtu.edu.cn)

**Abstract**

In this paper, a two-dimensional eight-velocity (D2Q8) multiple-relaxation-time (MRT) lattice Boltzmann (LB) model is proposed for incompressible porous flows at the representative elementary volume scale based on the Brinkman-Forchheimer-extended Darcy formulation. In the model, the porosity is included into the pressure-based equilibrium moments, and the linear and nonlinear drag forces of the porous media are incorporated into the model by adding a forcing term to the MRT-LB equation in the moment space. Through the Chapman-Enskog analysis, the generalized Navier-Stokes equations can be recovered exactly without artificial compressible errors. Numerical simulations of several typical two-dimensional porous flows are carried out to validate the present MRT-LB model. The numerical results of the present MRT-LB model are in good agreement with the analytical solutions and/or other numerical solutions reported in the literature.

## 1. Introduction

Fluid flow in porous media has gained significant research interest due to its importance of related scientific and industrial applications, which include contaminant transport in groundwater, crude oil exploration and extraction, radioactive waste management, hydrogeology and so on. Comprehensive literature surveys of these applications can be found in Refs. [1-3]. For incompressible flows in porous media at the representative elementary volume (REV) scale, the Darcy model, the Brinkman-extended Darcy model and the Forchheimer-extended Darcy model have been widely employed in many studies. However, the Darcy model and the two extended (Brinkman and Forchheimer) models have some intrinsic limitations in simulating porous flows [4]. In order to overcome the shortcomings of the above mentioned models, the Brinkman-Forchheimer-extended Darcy model (also called the generalized model) has been developed by several research groups [4-7]. In the generalized model, the viscous and inertial forces are incorporated into the momentum equation by using the local volume-averaging technique. The Darcy model and the two extended models can be regarded as the limiting cases of the generalized model. Moreover, the single phase flow and the transient flow in porous media can be solved by the generalized model [7]. As pointed out by Vafai and Kim [8], numerical results based on the Brinkman-Forchheimer-extended Darcy formulation have been shown to be in good agreement with the experimental predictions, and a porous medium/free-fluid interface can be best dealt with by the Brinkman-Forchheimer-extended Darcy formulation and the continuity of stresses and velocities at the interface. In the past several decades, various traditional numerical methods, such as the finite volume (FV) method, the finite difference (FD) method, and the finite element (FE) method, have been employed to study porous flows based on the generalized model.

The lattice Boltzmann (LB) method, as a mesoscopic numerical scheme originates from the lattice-gas automata (LGA) method [9], has achieved significant success in modeling complex fluid flows and related transport phenomena due to its kinetic characteristics [10-16]. Owing to the kinetic background, the LB method has some distinctive advantages over the traditional numerical methods (e.g., see Ref. [17]). Recently, the LB method has been successfully applied to simulate fluid flows in porous media at the REV scale [18-25]. In the REV scale method, the detailed geometric structure of the media is ignored, and the standard LB equation is modified by adding an additional term to account for the presence of the porous media. Therefore, LB method at the REV scale can be used for systems with large computational domain. It is worth mentioning that the REV LB method is very effective for simulating fluid flows in the region which is partially filled with a porous medium. As reported in Ref. [21], the discontinuity of the velocity-gradient at the porous medium/free-fluid interface can be well captured by the LB method without including the stress boundary condition into the simulation model.

However, to the best of our knowledge, most of the existing REV LB models [18-25] for incompressible porous flows employ the Bhatnagar-Gross-Krook (BGK) model (also called the singe-relaxation-time model) [26] to represent the collision process. Although the BGK model has become the most popular one because of its simplicity, there are several well-known criticisms on this model, such as the numerical instability at low values of viscosity [27-29] and the inaccuracy in treating boundary conditions [30]. On the other hand, it has been demonstrated that the deficiencies of the BGK model can be addressed by employing the multiple-relaxation-time (MRT) model [31]. Hence, the aim of this paper is to develop a new MRT-LB model for incompressible porous flows at the REV scale based on the generalized model. In the model, a pressure-based MRT-LB equation with the eight-by-eight collision matrix [32] is constructed in the framework of the standard MRT-LB method.

The remainder of this paper is organized as follows. In Section 2, the MRT-LB model for incompressible porous flows at the REV scale is presented. In Section 3, numerical tests of the

MRT-LB model are performed for the porous Poiseuille flow, porous Couette flow, lid-driven flow in a square porous cavity, and natural convection flow in a square porous cavity. Finally, a brief conclusion is made in Section 4.

## 2. MRT-LB model for incompressible flows in porous media

### 2.1 Macroscopic governing equations

The fluid flow is assumed to be two-dimensional, laminar and incompressible. For isothermal incompressible porous flows at the REV scale, the generalized model proposed by Nithiarasu *et al.* [4, 7] is employed in the present study. The dimensional governing equations of the generalized model can be written as

$$\nabla \cdot \mathbf{u} = 0, \tag{1}$$

$$\frac{\partial \mathbf{u}}{\partial t} + (\mathbf{u} \cdot \nabla)\left(\frac{\mathbf{u}}{\phi}\right) = -\frac{1}{\rho_0}\nabla(\phi p) + \upsilon_e \nabla^2 \mathbf{u} + \mathbf{F}, \tag{2}$$

where $\rho_0$ is the average fluid density, $\mathbf{u}$ and $p$ are the volume-averaged fluid velocity and pressure, respectively, $\phi$ is the porosity, and $\upsilon_e$ is the effective kinetic viscosity. $\mathbf{F} = (F_x, F_y)$ denotes the total body force induced by the porous matrix and other external forces, which can be expressed as [6, 22]

$$\mathbf{F} = -\frac{\phi \upsilon}{K}\mathbf{u} - \frac{\phi F_\phi}{\sqrt{K}}|\mathbf{u}|\mathbf{u} + \phi \mathbf{a}, \tag{3}$$

where $\upsilon$ is the kinetic viscosity of the fluid, $K$ is the permeability, $F_\phi$ is the geometric function, $\mathbf{a}$ is the body force due to an external force, and $|\mathbf{u}| = \sqrt{u_x^2 + u_y^2}$, in which $u_x$ and $u_y$ are the *x*- and *y*-components of the macroscopic velocity $\mathbf{u}$, respectively. Based on Ergun's relation [33], the geometric function $F_\phi$ and the permeability $K$ of the porous media can be expressed as [34]

$$F_\phi = \frac{1.75}{\sqrt{150\phi^3}}, \quad K = \frac{\phi^3 d_p^2}{150(1-\phi)^2}, \tag{4}$$

where $d_p$ is the solid particle diameter. The flow governed by Eqs. (1) and (2) are characterized by the porosity $\phi$ and several dimensionless parameters: the Darcy number $\mathrm{Da}$, the viscosity ratio $\mathrm{J}$, and the Reynolds number $\mathrm{Re}$, which are defined as

$$\mathrm{Da} = \frac{K}{L^2}, \quad \mathrm{J} = \frac{\upsilon_e}{\upsilon}, \quad \mathrm{Re} = \frac{LU}{\upsilon}, \tag{5}$$

where $L$ is the characteristic length, and $U$ is the characteristic velocity.

2.2 MRT-LB model

In this subsection, a two-dimensional MRT-LB model with eight velocities (D2Q8 model) is presented to study incompressible porous flows. According to Refs. [35, 36], the MRT-LB equation with an explicit treatment of the forcing term can be written as

$$\mathbf{f}(\mathbf{x}+\mathbf{e}\delta_t, t+\delta_t) - \mathbf{f}(\mathbf{x},t) = -\mathbf{M}^{-1}\mathbf{\Lambda}\left[\mathbf{m}(\mathbf{x},t) - \mathbf{m}^{(eq)}(\mathbf{x},t)\right] + \delta_t \mathbf{M}^{-1}\left(\mathbf{I} - \frac{\mathbf{\Lambda}}{2}\right)\mathbf{S}, \tag{6}$$

where $\mathbf{M}$ is the $8\times8$ orthogonal transformation matrix, $\mathbf{\Lambda}$ is the non-negative $8\times8$ diagonal relaxation matrix, and $\mathbf{I}$ is the $8\times8$ unit matrix. The boldface symbols, $\mathbf{f}$, $\mathbf{m}$, $\mathbf{m}^{(eq)}$, and $\mathbf{S}$ represent 8-dimensional (column) vectors:

$$\mathbf{f}(\mathbf{x},t) = \left(f_1(\mathbf{x},t), f_2(\mathbf{x},t), \cdots, f_8(\mathbf{x},t)\right)^\mathsf{T},$$

$$\mathbf{f}(\mathbf{x}+\mathbf{e}\delta_t, t+\delta_t) = \left(f_1(\mathbf{x}+\mathbf{e}_1\delta_t, t+\delta_t), \cdots, f_8(\mathbf{x}+\mathbf{e}_8\delta_t, t+\delta_t)\right)^\mathsf{T},$$

$$\mathbf{m}(\mathbf{x},t) = \left(m_1(\mathbf{x},t), m_2(\mathbf{x},t), \cdots, m_8(\mathbf{x},t)\right)^\mathsf{T},$$

$$\mathbf{m}^{(eq)}(\mathbf{x},t) = \left(m_1^{(eq)}(\mathbf{x},t), m_2^{(eq)}(\mathbf{x},t), \cdots, m_8^{(eq)}(\mathbf{x},t)\right)^\mathsf{T},$$

$$\mathbf{S} = \left(S_1, S_2, \cdots, S_8\right)^\mathsf{T},$$

where $\mathsf{T}$ is the transpose operator, $f_i(\mathbf{x},t)$ is the discrete distribution function corresponding to the discrete velocity $\mathbf{e}_i$ and time $t$, $\mathbf{m}(\mathbf{x},t)$ and $\mathbf{m}^{(eq)}(\mathbf{x},t)$ are the velocity moments of the discrete distribution functions $\mathbf{f}$ and the corresponding equilibrium moments, respectively, and $\{S_i \mid i=1, 2, \cdots, 8\}$ are components of the forcing term $\mathbf{S}$.

In the D2Q8 model, the eight discrete velocities $\{\mathbf{e}_i \,|\, i = 1, 2, \cdots, 8\}$ (see Fig. 1) are defined as [32, 37]

$$\mathbf{e}_i = \begin{cases} \left(\cos\left[(i-1)\pi/2\right], \sin\left[(i-1)\pi/2\right]\right)c, & i = 1 \sim 4 \\ \left(\cos\left[(2i-9)\pi/4\right], \sin\left[(2i-9)\pi/4\right]\right)\sqrt{2}c, & i = 5 \sim 8 \end{cases}, \quad (7)$$

where $c = \delta_x/\delta_t$, in which $\delta_t$ and $\delta_x$ are the time step and lattice spacing, respectively. The sound speed of the D2Q8 model is $c_s = c/\sqrt{3}$. In the present work, $c$ is set to be $1$, which leads to $\delta_x = \delta_t$.

The transformation matrix $\mathbf{M}$ linearly transforms the discrete distribution functions $\mathbf{f} \in \mathbb{V} = \mathbb{R}^8$ (velocity space) to their velocity moments $\mathbf{m} \in \mathbb{M} = \mathbb{R}^8$ (moment space):

$$\mathbf{m} = \mathbf{M}\mathbf{f}, \quad \mathbf{f} = \mathbf{M}^{-1}\mathbf{m}. \quad (8)$$

For the D2Q8 model, the eight velocity moments $\{m_i \,|\, i = 1, 2, \cdots, 8\}$ corresponding to the discrete velocities are:

$$\begin{aligned} \mathbf{m} &= \left(m_1, m_2, m_3, m_4, m_5, m_6, m_7, m_8\right)^{\mathrm{T}} \\ &= \left(p, e, j_x, q_x, j_y, q_y, p_{xx}, p_{xy}\right)^{\mathrm{T}}, \end{aligned} \quad (9)$$

where $m_1 = p$ is the pressure, $m_2 = e$ is related to energy, $m_{3,5} = j_{x,y}$ are components of the momentum $\mathbf{J} = (j_x, j_y)$, $m_{4,6} = q_{x,y}$ are related to energy flux, and $m_{7,8} = p_{xx,xy}$ are related to the diagonal and off-diagonal components of the stress tensor. With the ordering of the velocity moments $\{m_i\}$ given above, the transformation matrix $\mathbf{M}$ can be easily constructed ($c = 1$) [32]:

$$\mathbf{M} = \begin{pmatrix} 1 & 1 & 1 & 1 & 1 & 1 & 1 & 1 \\ -1 & -1 & -1 & -1 & 1 & 1 & 1 & 1 \\ 1 & 0 & -1 & 0 & 1 & -1 & -1 & 1 \\ -2 & 0 & 2 & 0 & 1 & -1 & -1 & 1 \\ 0 & 1 & 0 & -1 & 1 & 1 & -1 & -1 \\ 0 & -2 & 0 & 2 & 1 & 1 & -1 & -1 \\ 1 & -1 & 1 & -1 & 0 & 0 & 0 & 0 \\ 0 & 0 & 0 & 0 & 1 & -1 & 1 & -1 \end{pmatrix}. \quad (10)$$

The equilibrium moments $\mathbf{m}^{(\mathrm{eq})}$ for the velocity moments $\mathbf{m}$ are given as follows:

$$\mathbf{m}^{(eq)} = \left( \frac{5}{3}\phi p + \frac{2}{3}\frac{\rho_0 |\mathbf{u}|^2}{\phi}, e^{(eq)}, j_x, q_x^{(eq)}, j_y, q_y^{(eq)}, p_{xx}^{(eq)}, p_{xy}^{(eq)} \right)^{\mathrm{T}}, \tag{11}$$

where

$$e^{(eq)} = \alpha_2 \phi p + \frac{3}{4}\gamma_2 \frac{\rho_0 (u_x^2 + u_y^2)}{\phi},$$

$$q_x^{(eq)} = \frac{1}{2}c_1 \rho_0 u_x,$$

$$q_y^{(eq)} = \frac{1}{2}c_2 \rho_0 u_y,$$

$$p_{xx}^{(eq)} = \frac{3}{2}\gamma_1 \frac{\rho_0 (u_x^2 - u_y^2)}{\phi},$$

$$p_{xy}^{(eq)} = \frac{3}{2}\gamma_3 \frac{\rho_0 u_x u_y}{\phi}. \tag{12}$$

To get the correct generalized Navier-Stokes equations (1) and (2), the parameters are chosen as follows: $\alpha_2 = -1$, $c_1 = c_2 = -2$, $\gamma_2 = 0$, and $\gamma_1 = \gamma_3 = 2/3$. In the above equilibrium moments, we have employed the incompressibility approximation, i.e., the fluid density $\rho \approx \rho_0$ and $\mathbf{J} = (j_x, j_y) \approx \rho_0 \mathbf{u}$.

The relaxation matrix $\mathbf{\Lambda}$ is given by

$$\mathbf{\Lambda} = \mathrm{diag}(s_1, s_2, s_3, s_4, s_5, s_6, s_7, s_8)$$
$$= \mathrm{diag}(1, s_e, 1, s_q, 1, s_q, s_\upsilon, s_\upsilon). \tag{13}$$

The evolution process of the MRT-LB equation (6) consists of two steps: the collision step and streaming step [27]. Usually, the collision step is implemented in the moment space:

$$\mathbf{m}^+ = \mathbf{m} - \mathbf{\Lambda}\left[\mathbf{m} - \mathbf{m}^{(eq)}\right] + \delta_t \left(\mathbf{I} - \frac{\mathbf{\Lambda}}{2}\right)\mathbf{S}, \tag{14}$$

where $\mathbf{m}^+$ represent the velocity moments after collision, while the streaming step is still carried out in the velocity space:

$$f_i(\mathbf{x} + \mathbf{e}_i \delta_t, t + \delta_t) = f_i^+(\mathbf{x}, t), \tag{15}$$

where $\mathbf{f}^+ = \mathbf{M}^{-1}\mathbf{m}^+$.

In order to derive the correct equations of hydrodynamics, the forcing term $\mathbf{S}$ in the moment space should be chosen appropriately. For the D2Q8 MRT-LB model, the forcing term $\mathbf{S}$ can be chosen as

$$\mathbf{S} = \rho_0 \begin{pmatrix} 0 \\ 4\mathbf{u}\cdot\mathbf{F}/(5\phi) \\ F_x \\ -F_x \\ F_y \\ -F_y \\ 2(u_x F_x - u_y F_y)/\phi \\ (u_x F_y + u_y F_x)/\phi \end{pmatrix}, \tag{16}$$

where $\mathbf{F} = (F_x, F_y)$ is given by Eq. (3).

The fluid velocity $\mathbf{u}$ is defined as

$$\rho_0 \mathbf{u} = \sum_{i=1}^{8} \mathbf{e}_i f_i + \frac{\delta_t}{2}\rho_0 \mathbf{F}, \tag{17}$$

Note that the total body force $\mathbf{F}$ also contains the velocity $\mathbf{u}$, so Eq. (17) is a nonlinear equation for $\mathbf{u}$. According to Eqs (3) and (17), the velocity $\mathbf{u}$ can be calculated explicitly by

$$\mathbf{u} = \frac{\mathbf{v}}{l_0 + \sqrt{l_0^2 + l_1\,|\mathbf{v}|}}, \tag{18}$$

where $\mathbf{v}$, $l_0$ and $l_1$ are given by

$$\mathbf{v} = \sum_{i=1}^{8} \mathbf{e}_i f_i/\rho_0 + \frac{\delta_t}{2}\phi \mathbf{a}, \quad l_0 = \frac{1}{2}\left(1 + \phi\frac{\delta_t}{2}\frac{\upsilon}{K}\right), \quad l_1 = \phi\frac{\delta_t}{2}\frac{F_\phi}{\sqrt{K}}. \tag{19}$$

According to Ref. [39], the fluid pressure $p$ can be determined by

$$p = \frac{c_s^2}{\phi(1-\omega_0)}\left[\sum_{i=1}^{8} f_i + \rho_0 s_0(\mathbf{u})\right], \tag{20}$$

where $\omega_0 = 4/9$ and $s_0(\mathbf{u}) = -2\rho_0|\mathbf{u}|^2/(3\phi)$. Here, $s_i(\mathbf{u}) = \omega_i\left[3\mathbf{e}_i\cdot\mathbf{u} + 4.5(\mathbf{e}_i\cdot\mathbf{u})^2/\phi - 1.5|\mathbf{u}|^2/\phi\right]$ with $\omega_0 = 4/9$, $\omega_{1\sim4} = 1/9$ and $\omega_{5\sim8} = 1/36$.

The effective kinetic viscosity $\upsilon_e$ in the model is defined as

$$\upsilon_e = c_s^2 \left(\tau_\upsilon - \frac{1}{2}\right)\delta_t \tag{21}$$

with $s_7 = s_8 = s_\upsilon = 1/\tau_\upsilon$. Through the Chapman-Enskog analysis [32, 35, 38] of the MRT-LB equation (6) in the moment space, the generalized Navier-Stokes equations (1) and (2) can be recovered exactly in the incompressible limit (see Appendix A for details). If the eight relaxation rates $\{s_i | i = 1, 2, \cdots, 8\}$ are set to be a single value $1/\tau_\upsilon$, i.e., $\Lambda = (1/\tau_\upsilon)\mathbf{I}$, then the present MRT-LB model reduces to a BGK-LB model with the following equilibrium distribution function:

$$f_i^{(eq)} = \rho_0 \omega_i \left[\frac{\phi p}{\rho_0 c_s^2} + \frac{\mathbf{e}_i \cdot \mathbf{u}}{c_s^2} + \frac{(\mathbf{e}_i \cdot \mathbf{u})^2}{2\phi c_s^4} - \frac{|\mathbf{u}|^2}{2\phi c_s^2}\right]. \tag{22}$$

It should be noted that, as $\phi \to 1$ and $\text{Da} \to \infty$, the present MRT-LB model reduces to the incompressible MRT-LB model [32] for free-fluid flows without porous media. When $F_\phi = 0$, the Brinkman-extended Darcy equation can be obtained from the present MRT-LB model.

## 3. Numerical tests

In this section, numerical simulations of several typical two-dimensional porous flows are carried out to validate the proposed MRT-LB model. The testing problems include the Poiseuille flow in a channel filled with porous media, the Couette flow between two parallel plates filled with porous media, the lid-driven flow in a square porous cavity, and the natural convection flow in a square porous cavity. The present numerical results are compared with the analytical and numerical solutions in previous studies. In simulations, we set $\rho_0 = 1$, $c = 1$, and $\delta_t = \delta_x = \delta_y = 1$. The relaxation rates $\{s_i | 1 \leq i \leq 8\}$ are chosen as follows: $s_1 = s_3 = s_5 = 1$, $s_2 = 1.1$, $s_4 = s_6 = 1.2$, and $s_7 = s_8 = s_\upsilon = 1/\tau_\upsilon$. Unless otherwise stated, the non-equilibrium extrapolation scheme [40] is employed to treat the velocity and temperature boundary conditions in simulations.

### 3.1 Poiseuille flow in a channel filled with porous media

The first test problem is the Poiseuille flow in a 2D channel filled with porous media. The height of the channel is $L$, and the flow is driven by an external force $\mathbf{a} = (a_x, 0)$ along the channel direction. When the flow is fully developed in the *x*-direction (along the channel direction), the governing equation of the flow can be expressed as

$$\frac{\upsilon_e}{\phi} \frac{\partial^2 u_x}{\partial y^2} - \frac{\upsilon}{K} u_x - \frac{F_\phi}{\sqrt{K}} u_x^2 + a_x = 0 \tag{23}$$

with $u_x(x,0) = u_x(x,L) = 0$, and the *y*-direction velocity component $u_y$ is zero everywhere. The Reynolds number Re is defined as $\text{Re} = L u_0 / \upsilon$, where $u_0$ is the maximum velocity of the porous Poiseuille flow (without the nonlinear drag force, i.e., $F_\phi = 0$) along the centerline of the channel [22]:

$$u_0 = \frac{a_x K}{\upsilon} \left[ 1 - \cosh^{-1}\left(\frac{\vartheta L}{2}\right) \right], \tag{24}$$

where $\vartheta = \sqrt{\phi \upsilon / (K \upsilon_e)}$.

In simulations, the porosity $\phi$ is set to be 0.1, the viscosity ratio $J$ is set to be 1, the Darcy number Da changes from $10^{-6}$ to $10^{-3}$, and the Reynolds number Re changes from 0.01 to 100. The relaxation rate $s_\upsilon$ is set to be $5/3$ ($\tau_\upsilon = 0.6$) with a $N_x \times N_y = 80 \times 80$ square mesh. The nonequilibrium extrapolation scheme is employed to the top and bottom plates for no-slip velocity boundary condition, and the periodic boundary conditions are employed to the inlet and outlet of the channel. At $t = 0$, the velocity moments are set to be their equilibrium values, and the velocity field is initialized to be 0 with a constant pressure $p = 1$. The velocity profiles predicted by the present MRT-LB model for different Re and Da are shown in Fig. 2. The numerical solutions given by Guo and Zhao [22] using the FD method are also included in the figure for comparison. From Fig. 2 it is clearly seen that the present numerical results agree well with the FD solutions reported in the literature.

**3.2 Couette flow between two parallel plates filled with porous media**

We now apply the MRT-LB model to the Couette flow in a 2D channel filled with fluid-saturated porous media. The flow is driven by the top plate moving along the *x*-direction with a uniform velocity $u_0$, while the bottom plate is fixed. The Reynolds number Re of this flow is defined by $\text{Re} = Lu_0/\upsilon$. At the steady state, the flow still follows Eq. (24) ($a_x = 0$) with $u_x(x,0) = 0$ and $u_x(x,L) = u_0$. In simulations, the porosity $\phi$ is set to be 0.1, the viscosity ratio J is set to be 1, and the boundary and initial conditions are the same as those in the porous Poiseuille flow. For Re = 0.001, 10, 50, and 1000, the kinetic viscosity $\upsilon$ is set to be 0.4, 0.08, 0.032, and 0.0008 based on a $N_x \times N_y = 80 \times 80$ square mesh, respectively. In Fig. 3, the velocity profiles predicted by the present MRT-LB model are presented for various Re and Da. The FD solutions in Ref. [22] are also included in Fig. 3 for comparison. Excellent agreement can be found between the present MRT-LB results and the FD solutions.

The MRT-LB model is also applied to simulate the modified Couette flow in a 2D channel partially filled with a fluid-saturated porous medium, which has been investigated in Refs. [21, 22, 41]. A porous layer is positioned in the lower part of the channel such that there is a gap between the top plate and the medium. The porosity $\phi$ is set to be 0.1 for $0 \leq y/L \leq 1/2$ (porous region) and 1 for $1/2 < y/L \leq 1$ (free-fluid region). For small Re and Da, the nonlinear drag force can be neglected, and the Brinkman model is applied in the porous region. At the steady state, the analytical solution of the flow can be expressed as [22, 41]:

$$u_x(y) = \begin{cases} d_0 + d_1(y/L - 1/2) & 1/2 < y/L \leq 1 \\ d_0 \exp[\vartheta(y/L - 1/2)] & 0 \leq y/L \leq 1/2 \end{cases}, \quad (25)$$

where $\vartheta = \sqrt{\phi\upsilon/(K\upsilon_e)}$, $d_0 = 2\vartheta K u_0/(2\vartheta K + \phi)$, and $d_1 = 2\phi u_0/(2\vartheta K + \phi)$. In simulations, $F_\phi$ is set to be zero, and the fluid kinetic viscosity $\upsilon$ is set to be 0.16 with a $N_x \times N_y = 80 \times 80$ lattice. The relaxation rate $s_\upsilon$ is determined by $s_\upsilon = 1/\left[0.5 + J\upsilon/(c_s^2 \delta_t)\right]$ (for free-fluid region, J = 1). The velocity profiles predicted by the present MRT-LB model for $J=1$ and $4$ at $Da=0.001$ and $Re=0.01$ are plotted in Fig. 4. As shown, the present numerical results agree well with the analytical ones. The discontinuity of the velocity-gradient at the porous medium/free-fluid interface is well captured by the present MRT-LB model without any special treatment for the boundary condition at the interface in simulations.

**3.3 Lid-driven flow in a square porous cavity**

In this subsection, we apply the MRT-LB model to simulate the incompressible flow in a square cavity filled with a fluid-saturated porous medium. The top wall of the cavity moves from left to right with a uniform velocity $u_0$, while the bottom, right, and left walls of the cavity are fixed. The Reynolds number $Re$ of this flow is defined as $Re=Lu_0/\upsilon$, where $L$ is the height of the cavity. In simulations, we set $\phi=0.1$, $Re=10$, $J=1$, and the kinetic viscosity $\upsilon$ is set to be $0.256$ based on a $N_x \times N_y = 128 \times 128$ square mesh. The nonequilibrium extrapolation scheme is adopted to treat the velocity boundary conditions of $\{f_i\}$ on the four walls. The velocity moments are set to be their equilibrium values, and the velocity field is initialized to be $0$ with a constant pressure $p=1$ at $t=0$. The streamlines predicted by the present MRT-LB model for different Darcy numbers with $\phi=0.1$ and $Re=10$ are illustrated in Fig. 5. From the figure we can see that, as $Da$ decreases, the vortex inside the cavity becomes weaker, and the boundary layer near the top wall becomes thinner. This can be attributed to the lower permeability ($K=DaL^2$) of the medium which results in lower fluid velocity. To be more informative, the horizontal velocity component $u_x$ in the vertical midplane $x=L/2$ and the vertical velocity component $u_y$ in the horizontal midplane $y=L/2$ are plotted in Fig. 6. The FD solutions in Ref. [22] are also plotted in Fig. 6 for comparison. Obviously, the present numerical results are in good agreement with the FD solutions.

As $\phi \to 1$ and Da tends to infinity, the present MRT-LB model reduces to the incompressible MRT-LB model for free-fluid flows without porous media. We now apply the MRT-LB model to the lid-driven cavity flow without porous media. In simulations, we set $\phi = 0.999$, $\text{Da} = 10^6$, and $u_0 = 0.1$. In Fig. 7, the velocity profiles through the center of the cavity are plotted for different Reynolds numbers. The benchmark solutions of Ghia *et al*. [42] are also presented for comparison. Excellent agreement can be found between the present MRT-LB results and the benchmark solutions.

**3.4 Natural convection flow in a square cavity filled with porous media**

Natural convection flow in a square cavity filled with a porous medium (see Fig. 8) has been studied extensively by many researchers [4, 7, 43, 44] based on the generalized model. For this problem, the velocity field is solved by the present MRT-LB model and the temperature field is solved by the D2Q5 MRT-LB model [44] (see Appendix B for details). The top and bottom walls of the cavity are thermally insulated, while the left and right vertical walls are kept at constant but different temperatures $T_h$ and $T_c$ ($T_h > T_c$), respectively. The Prandtl number Pr and Rayleigh number Ra of this flow are defined as $\text{Pr} = \upsilon/\alpha$ and $\text{Ra} = g\beta\Delta T L^3 \text{Pr}/\upsilon^2$, respectively, where $\Delta T = T_h - T_c$ is the temperature difference, $\alpha$ is the thermal diffusivity of the fluid, $g$ is the gravitational acceleration, $\beta$ is the thermal expansion coefficient, and $L$ is the distance between the walls. The body force **a** is defined as $\mathbf{a} = g\beta(T - T_0)\mathbf{j}$, where $T_0 = (T_h + T_c)/2$ is the reference temperature, and **j** is the unit vector in the *y*-direction.

In simulations, we set $\text{Pr} = 1$, $\text{J} = 1$, $\sigma = 1$ (thermal capacity ratio), $\gamma = \alpha_e/\alpha = 1$ ($\alpha_e$ is the effective thermal diffusivity), $\zeta_0 = 1$, $\zeta_1 = \zeta_2 = 1/\tau_T$, $\zeta_3 = \zeta_4 = 1.1$, $T_h = 21$, and $T_c = 1$. According to Refs. [13, 45], the dimensionless relaxation times of the velocity and temperature fields can be determined as

$$\tau_\upsilon = \frac{1}{2} + \frac{\text{Ma}JL\sqrt{3\text{Pr}}}{c\delta_t\sqrt{\text{Ra}}}, \quad \tau_T = \frac{1}{2} + \frac{\gamma c_s^2(\tau_\upsilon - 0.5)}{J\sigma c_{sT}^2 \text{Pr}}, \tag{26}$$

respectively, where $\text{Ma} = U/c_s = \sqrt{3}U$ is the Mach number ($U = \sqrt{\beta g \Delta TL}$ is the characteristic velocity, and usually $\text{Ma} \leq 0.3$), and $c_{sT}^2 = 2c^2/5$ ($c_{sT}$ is the sound speed of the D2Q5 model). The grid sizes of $120 \times 120$, $200 \times 200$, and $250 \times 250$ are employed for $\text{Da} = 10^{-2}$, $10^{-4}$, and $10^{-6}$, respectively. Fig. 9 illustrates the streamlines and isotherms for different Rayleigh numbers and Darcy numbers (Darcy-Rayleigh number $\text{Ra}^* = \text{Ra}\,\text{Da} = 1000$) with $\text{Pr} = 1$ and $\phi = 0.6$. From the figure we can observe that, as Da decreases, the velocity and thermal boundary layers become thinner near the hot and cold vertical walls. As Da increases to $10^{-2}$, the streamlines and isotherms are less crowded near the vertical walls and more convective mixing occurs inside the cavity. To quantify the results, the average Nusselt numbers of the left vertical wall are calculated and listed in Table 1. The numerical results given by Nithiarasu *et al*. [4, 7] using the finite element method are also included in Table 1 for comparison. To sum up, the numerical results of the present MRT-LB model agree well with those results reported in previous studies.

## 4. Conclusion

In this paper, an MRT-LB model with the eight-by-eight collision matrix has been presented for simulating incompressible flows in porous media at the REV scale. The key point of the MRT-LB model is to include the porosity into the equilibrium moments, and add a forcing term to the MRT-LB equation in the moment space to account for the Darcy (linear) and Forchheimer (nonlinear) drag forces of the solid matrix based on the generalized model. Through the Chapman-Enskog analysis in the moment space, the generalized Navier-Stokes equations can be recovered exactly without artificial compressible errors. Numerical simulations of the porous Poiseuille flow, porous Couette flow, lid-driven flow in square porous cavity, and natural convection flow in a square porous cavity have been carried out to demonstrate the present MRT-LB model. The numerical results of the present MRT-LB model agree well with the analytical solutions and/or other numerical solutions reported in previous studies.

**Acknowledgements**

This work was supported by the National Key Basic Research Program of China (973 Program) (2013CB228304).

**Appendix A: Chapman-Enskog analysis of the D2Q8 MRT-LB model**

The Chapman-Enskog expansion method [32, 35, 38] is adopted to derive the generalized Navier-Stokes equations (1) and (2) from the D2Q8 MRT-LB model. To this end, the following expansions in time and space are introduced:

$$f_i(\mathbf{x}+\mathbf{e}_i\delta_t,\ t+\delta_t) = \sum_{i=0}^{\infty}\frac{\epsilon^n}{n!}(\partial_t+\mathbf{e}_i\cdot\nabla)^n f_i(\mathbf{x},\ t), \quad \text{(A.1a)}$$

$$f_i = f_i^{(0)} + \epsilon f_i^{(1)} + \epsilon^2 f_i^{(2)} + \cdots, \quad \text{(A.1b)}$$

$$\partial_t = \epsilon\partial_{t_1} + \epsilon^2\partial_{t_2},\ \partial_j = \epsilon\partial_{j1},\ \mathbf{S} = \epsilon\mathbf{S}_1,\ \mathbf{F} = \epsilon\mathbf{F}_1, \quad \text{(A.1c)}$$

where $\epsilon = \delta_t$ is a small expansion parameter, $\mathbf{S}_1 = (S_{11},\ldots,S_{81})^\mathrm{T}$, $\mathbf{F}_1 = (F_{x1}, F_{y1})$. With the above expansions, we can derive the following equations from Eq. (6) as consecutive orders of the parameter $\epsilon$ in the moment space as

$$\epsilon^0: \quad \mathbf{m}^{(0)} = \mathbf{m}^{(eq)}, \tag{A.2a}$$

$$\epsilon^1: \quad \tilde{\mathbf{D}}_1 \mathbf{m}^{(0)} = -\mathbf{\Lambda}' \mathbf{m}^{(1)} + \left(\mathbf{I} - \frac{\mathbf{\Lambda}}{2}\right)\mathbf{S}_1, \tag{A.2b}$$

$$\epsilon^2: \quad \partial_{t_2} \mathbf{m}^{(0)} + \tilde{\mathbf{D}}_1 \left(\mathbf{I} - \frac{\mathbf{\Lambda}}{2}\right)\mathbf{m}^{(1)} + \frac{\delta_t}{2} \tilde{\mathbf{D}}_1 \left(\mathbf{I} - \frac{\mathbf{\Lambda}}{2}\right)\mathbf{S}_1 = -\mathbf{\Lambda}' \mathbf{m}^{(2)}, \tag{A.2c}$$

where $\tilde{\mathbf{D}}_1 = \mathbf{M}\mathbf{D}_1\mathbf{M}^{-1} = \partial_{t_1}\mathbf{I} + \mathbf{C}_j \partial_{j1}$ ($j = x, y$), $\mathbf{D}_1 = \partial_{t_1}\mathbf{I} + \partial_{j1}\mathrm{diag}(e_{1j}, e_{2j}, \cdots, e_{8j})$, $\mathbf{C}_j = \mathbf{M}(e_{ij}\mathbf{I})\mathbf{M}^{-1}$, $\mathbf{\Lambda}' = \mathbf{\Lambda}/\delta_t$, and

$$\mathbf{m}^{(1)} = \left(0, e^{(1)}, -\delta_t \rho_0 F_{x1}/2, q_x^{(1)}, -\delta_t \rho_0 F_{y1}/2, q_y^{(1)}, p_{xx}^{(1)}, p_{xy}^{(1)}\right)^{\mathrm{T}}. \tag{A.3}$$

$\mathbf{C}_x$ and $\mathbf{C}_y$ can be given explicitly by

$$\mathbf{C}_x = \begin{pmatrix} 0 & 0 & 1 & 0 & 0 & 0 & 0 & 0 \\ 0 & 0 & \frac{1}{3} & \frac{2}{3} & 0 & 0 & 0 & 0 \\ \frac{3}{4} & \frac{1}{4} & 0 & 0 & 0 & 0 & \frac{1}{2} & 0 \\ 0 & 1 & 0 & 0 & 0 & 0 & -1 & 0 \\ 0 & 0 & 0 & 0 & 0 & 0 & 0 & 1 \\ 0 & 0 & 0 & 0 & 0 & 0 & 0 & 1 \\ 0 & 0 & \frac{1}{3} & -\frac{1}{3} & 0 & 0 & 0 & 0 \\ 0 & 0 & 0 & 0 & \frac{2}{3} & \frac{1}{3} & 0 & 0 \end{pmatrix}, \quad \mathbf{C}_y = \begin{pmatrix} 0 & 0 & 0 & 0 & 1 & 0 & 0 & 0 \\ 0 & 0 & 0 & 0 & \frac{1}{3} & \frac{2}{3} & 0 & 0 \\ 0 & 0 & 0 & 0 & 0 & 0 & 0 & 1 \\ 0 & 0 & 0 & 0 & 0 & 0 & 0 & 1 \\ \frac{3}{4} & \frac{1}{4} & 0 & 0 & 0 & 0 & -\frac{1}{2} & 0 \\ 0 & 1 & 0 & 0 & 0 & 0 & 1 & 0 \\ 0 & 0 & 0 & 0 & -\frac{1}{3} & \frac{1}{3} & 0 & 0 \\ 0 & 0 & \frac{2}{3} & \frac{1}{3} & 0 & 0 & 0 & 0 \end{pmatrix}. \tag{A.4}$$

From Eq. (A.2b), the following equations at the $t_1$ time scale can be obtained:

$$\partial_{t_1}\left(\frac{5}{3}\phi p + \frac{2}{3}\frac{\rho_0 |\mathbf{u}|^2}{\phi}\right) + \partial_{x1}(\rho_0 u_x) + \partial_{y1}(\rho_0 u_y) = 0, \tag{A.5a}$$

$$\partial_{t_1}(-\phi p) - \frac{1}{3}\partial_{x1}(\rho_0 u_x) - \frac{1}{3}\partial_{y1}(\rho_0 u_y) = -s_2' e^{(1)} + \left(1 - \frac{s_2}{2}\right)S_{21}, \tag{A.5b}$$

$$\partial_{t_1}(\rho_0 u_x) + \partial_{x1}\left(\phi p + \frac{\rho_0 u_x^2}{\phi}\right) + \partial_{y1}\left(\frac{\rho_0 u_x u_y}{\phi}\right) = \frac{\delta_t}{2}s_3' \rho_0 F_{x1} + \left(1 - \frac{s_3}{2}\right)S_{31}, \tag{A.5c}$$

$$\partial_{t_1}(-\rho_0 u_x) + \partial_{x1}\left[-\phi p - \frac{\rho_0(u_x^2 - u_y^2)}{\phi}\right] + \partial_{y1}\left(\frac{\rho_0 u_x u_y}{\phi}\right) = -s_4' q_x^{(1)} + \left(1 - \frac{s_4}{2}\right)S_{41}, \tag{A.5d}$$

$$\partial_{t_1}(\rho_0 u_y) + \partial_{x1}\left(\frac{\rho_0 u_x u_y}{\phi}\right) + \partial_{y1}\left(\phi p + \frac{\rho_0 u_y^2}{\phi}\right) = \frac{\delta_t}{2}s_5' \rho_0 F_{y1} + \left(1 - \frac{s_5}{2}\right)S_{51}, \tag{A.5e}$$

$$\partial_{t_1}(-\rho_0 u_y) + \partial_{x1}\left(\frac{\rho_0 u_x u_y}{\phi}\right) + \partial_{y1}\left[-\phi p + \frac{\rho_0(u_x^2 - u_y^2)}{\phi}\right] = -s_6' q_y^{(1)} + \left(1 - \frac{s_6}{2}\right)S_{61}, \tag{A.5f}$$

$$\partial_{t_1}\left[\frac{\rho_0(u_x^2 - u_y^2)}{\phi}\right] + \frac{2}{3}\partial_{x1}(\rho_0 u_x) - \frac{2}{3}\partial_{y1}(\rho_0 u_y) = -s_7' p_{xx}^{(1)} + \left(1 - \frac{s_7}{2}\right)S_{71}, \tag{A.5g}$$

$$\partial_{t_1}\left(\frac{\rho_0 u_x u_y}{\phi}\right)+\frac{1}{3}\partial_{x1}(\rho_0 u_y)+\frac{1}{3}\partial_{y1}(\rho_0 u_x)=-s_8' p_{xy}^{(1)}+\left(1-\frac{s_8}{2}\right)S_{81}. \tag{A.5h}$$

From Eq. (A.2c), the following equations at the $t_2$ time scale can be obtained:

$$\partial_{t_2}\left(\frac{5}{3}\phi p+\frac{2}{3}\frac{\rho_0 |\mathbf{u}|^2}{\phi}\right)=0, \tag{A.6a}$$

$$\partial_{t_2}(\rho_0 u_x)-\frac{\delta_t}{2}\partial_{t_1}\left[\left(1-\frac{s_3}{2}\right)\rho_0 F_{x1}\right]+\partial_{x1}\left[\frac{1}{4}\left(1-\frac{s_2}{2}\right)e^{(1)}+\frac{1}{2}\left(1-\frac{s_7}{2}\right)p_{xx}^{(1)}\right]+\partial_{y1}\left[\left(1-\frac{s_8}{2}\right)p_{xy}^{(1)}\right]$$

$$+\frac{\delta_t}{2}\left\{\partial_{t_1}\left[\left(1-\frac{s_3}{2}\right)S_{31}\right]+\partial_{x1}\left[\frac{1}{4}\left(1-\frac{s_2}{2}\right)S_{21}+\frac{1}{2}\left(1-\frac{s_7}{2}\right)S_{71}\right]+\partial_{y1}\left[\left(1-\frac{s_8}{2}\right)S_{81}\right]\right\}=0, \tag{A.6b}$$

$$\partial_{t_2}(\rho_0 u_y)-\frac{\delta_t}{2}\partial_{t_1}\left[\left(1-\frac{s_5}{2}\right)\rho_0 F_{y1}\right]+\partial_{x1}\left[\left(1-\frac{s_8}{2}\right)p_{xy}^{(1)}\right]+\partial_{y1}\left[\frac{1}{4}\left(1-\frac{s_2}{2}\right)e^{(1)}-\frac{1}{2}\left(1-\frac{s_7}{2}\right)p_{xx}^{(1)}\right]$$

$$+\frac{\delta_t}{2}\left\{\partial_{t_1}\left[\left(1-\frac{s_5}{2}\right)S_{51}\right]+\partial_{x1}\left[\left(1-\frac{s_8}{2}\right)S_{81}\right]+\partial_{y1}\left[\frac{1}{4}\left(1-\frac{s_2}{2}\right)S_{21}-\frac{1}{2}\left(1-\frac{s_7}{2}\right)S_{71}\right]\right\}=0, \tag{A.6c}$$

Following the approach in Ref. [32], we can obtain:

$$\partial_{t_1}\left(\frac{5}{3}\phi p+\frac{2}{3}\frac{\rho_0 |\mathbf{u}|^2}{\phi}\right)=0, \tag{A.7}$$

Combining Eqs. (A.5a) and (A.6a) with Eq. (A.7) leads to the following incompressible continuity equation

$$\partial_x u_x+\partial_y u_y=0, \tag{A.8}$$

Note that $e^{(1)}$, $p_{xx}^{(1)}$ and $p_{xy}^{(1)}$ in Eqs. (A.6b) and (A.6c) are unknowns to be determined. According to Eqs. (A.5b), (A.5g) and (A.5h), we can get:

$$-s_2' e^{(1)}=\partial_{t_1}(-\phi p)-\frac{1}{3}\partial_{x1}(\rho_0 u_x)-\frac{1}{3}\partial_{y1}(\rho_0 u_y)-\left(1-\frac{s_2}{2}\right)S_{21}, \tag{A.9a}$$

$$-s_7' p_{xx}^{(1)}=\partial_{t_1}\left[\frac{\rho_0(u_x^2-u_y^2)}{\phi}\right]+\frac{2}{3}\partial_{x1}(\rho_0 u_x)-\frac{2}{3}\partial_{y1}(\rho_0 u_y)-\left(1-\frac{s_7}{2}\right)S_{71}, \tag{A.9b}$$

$$-s_8' p_{xy}^{(1)}=\partial_{t_1}\left(\frac{\rho_0 u_x u_y}{\phi}\right)+\frac{1}{3}\partial_{x1}(\rho_0 u_y)+\frac{1}{3}\partial_{y1}(\rho_0 u_x)-\left(1-\frac{s_8}{2}\right)S_{81}. \tag{A.9c}$$

Neglecting the terms of order $O(|\mathbf{u}|^3)$ and higher-order terms of the form $u_j \partial_k(u_k u_j)$, using Eqs. (A.5c) and (A.5e), we can obtain:

$$\partial_{t_1}\left(\frac{\rho_0 u_x^2}{\phi}\right) = \frac{2\rho_0 u_x F_{x1}}{\phi}, \tag{A.10a}$$

$$\partial_{t_1}\left(\frac{\rho_0 u_y^2}{\phi}\right) = \frac{2\rho_0 u_y F_{y1}}{\phi}, \tag{A.10b}$$

$$\partial_{t_1}\left(\frac{\rho_0 u_x u_y}{\phi}\right) = \frac{\rho_0\left(u_x F_{y1} + u_y F_{x1}\right)}{\phi}. \tag{A.10c}$$

With the above equations, we can obtain:

$$-s_2' e^{(1)} = \frac{4\rho_0}{15}\left(\partial_{x1} u_x + \partial_{y1} u_y\right) + \frac{2s_2}{5}\frac{\rho_0\left(u_x F_{x1} + u_y F_{y1}\right)}{\phi}, \tag{A.11a}$$

$$-s_7' p_{xx}^{(1)} = \frac{2\rho_0}{3}\left(\partial_{x1} u_x - \partial_{y1} u_y\right) + s_7 \frac{\rho_0\left(u_x F_{x1} - u_y F_{y1}\right)}{\phi}, \tag{A.11b}$$

$$-s_8' p_{xy}^{(1)} = \frac{\rho_0}{3}\left(\partial_{x1} u_y + \partial_{y1} u_x\right) + \frac{s_8}{2}\frac{\rho_0\left(u_x F_{y1} + u_y F_{x1}\right)}{\phi}. \tag{A.11c}$$

Substituting Eq. (A.11) into Eq. (A.6), the following equations at the $t_2$ time scale can be derived:

$$\partial_{t_2} u_x - \frac{\delta_t}{15}\left(\frac{1}{s_2} - \frac{1}{2}\right)\partial_{x1}\left(\partial_{x1} u_x + \partial_{y1} u_y\right) - \frac{\delta_t}{3}\left(\frac{1}{s_7} - \frac{1}{2}\right)\partial_{x1}\left(\partial_{x1} u_x - \partial_{y1} u_y\right)$$

$$-\frac{\delta_t}{3}\left(\frac{1}{s_8} - \frac{1}{2}\right)\partial_{y1}\left(\partial_{x1} u_y + \partial_{y1} u_x\right) = 0, \tag{A.12a}$$

$$\partial_{t_2} u_y - \frac{\delta_t}{3}\left(\frac{1}{s_8} - \frac{1}{2}\right)\partial_{x1}\left(\partial_{x1} u_y + \partial_{y1} u_x\right) - \frac{\delta_t}{15}\left(\frac{1}{s_2} - \frac{1}{2}\right)\partial_{y1}\left(\partial_{x1} u_x + \partial_{y1} u_y\right)$$

$$+\frac{\delta_t}{3}\left(\frac{1}{s_7} - \frac{1}{2}\right)\partial_{y1}\left(\partial_{x1} u_x - \partial_{y1} u_y\right) = 0. \tag{A.12b}$$

Combining Eq. (A.12) ($t_2$ time scale) with Eqs. (A.5c) and (A.5e) ($t_1$ time scale), the following equations can be derived ($\partial_t = \epsilon \partial_{t_1} + \epsilon^2 \partial_{t_2}$):

$$\partial_t u_x + \partial_x\left(\frac{u_x^2}{\phi}\right) + \partial_y\left(\frac{u_x u_y}{\phi}\right) = -\frac{1}{\rho_0}\partial_x(\phi p) + \frac{\delta_t}{15}\left(\frac{1}{s_2} - \frac{1}{2}\right)\partial_x\left(\partial_x u_x + \partial_y u_y\right)$$

$$+\frac{\delta_t}{3}\left(\frac{1}{s_7}-\frac{1}{2}\right)\partial_x\left(\partial_x u_x-\partial_y u_y\right)+\frac{\delta_t}{3}\left(\frac{1}{s_8}-\frac{1}{2}\right)\partial_y\left(\partial_x u_y+\partial_y u_x\right)+F_x, \quad \text{(A.13a)}$$

$$\partial_t u_y+\partial_x\left(\frac{u_x u_y}{\phi}\right)+\partial_y\left(\frac{u_y^2}{\phi}\right)=-\frac{1}{\rho_0}\partial_y(\phi p)+\frac{\delta_t}{3}\left(\frac{1}{s_8}-\frac{1}{2}\right)\partial_x\left(\partial_x u_y+\partial_y u_x\right)$$

$$+\frac{\delta_t}{15}\left(\frac{1}{s_2}-\frac{1}{2}\right)\partial_y\left(\partial_x u_x+\partial_y u_y\right)-\frac{\delta_t}{3}\left(\frac{1}{s_7}-\frac{1}{2}\right)\partial_y\left(\partial_x u_x-\partial_y u_y\right)+F_y. \quad \text{(A.13b)}$$

With the aid of the incompressible continuity equation $\partial_x u_x+\partial_y u_y=0$, the incompressible generalized Navier-Stokes equations (1) and (2) can be obtained from Eq. (A.13). The effective kinetic viscosity is defined as $\upsilon_e=c_s^2(\tau_\upsilon-0.5)\delta_t$ with $s_7=s_8=s_\upsilon=1/\tau_\upsilon$.

**Appendix B: D2Q5 MRT-LB model**

The macroscopic temperature equation of natural convection flow in porous media can be written as

$$\sigma\frac{\partial T}{\partial t}+\mathbf{u}\cdot\nabla T=\nabla\cdot(\alpha_e\nabla T), \quad \text{(B.1)}$$

where $\sigma$ is the thermal capacity ratio, and $\alpha_e$ is the effective thermal diffusivity. For the temperature field, the D2Q5 MRT-LB equation is defined as

$$\mathbf{g}(\mathbf{x}+\mathbf{e}\delta_t,t+\delta_t)-\mathbf{g}(\mathbf{x},t)=-\mathbf{N}^{-1}\Theta\left[\mathbf{n}(\mathbf{x},t)-\mathbf{n}^{(eq)}(\mathbf{x},t)\right], \quad \text{(B.2)}$$

where $\mathbf{N}$ is a $5\times5$ orthogonal transformation matrix, and $\Theta$ is a diagonal relaxation matrix. The boldface symbols, $\mathbf{g}$, $\mathbf{n}$, and $\mathbf{n}^{(eq)}$ are 5-dimensional (column) vectors:

$$\mathbf{g}(\mathbf{x},t)=\left(g_0(\mathbf{x},t),g_1(\mathbf{x},t),\cdots,g_4(\mathbf{x},t)\right)^T,$$

$$\mathbf{g}(\mathbf{x}+\mathbf{e}\delta_t,t+\delta_t)=\left(g_0(\mathbf{x}+\mathbf{e}_0\delta_t,t+\delta_t),\cdots,g_4(\mathbf{x}+\mathbf{e}_4\delta_t,t+\delta_t)\right)^T,$$

$$\mathbf{n}(\mathbf{x},t)=\left(n_0(\mathbf{x},t),n_1(\mathbf{x},t),\cdots,n_4(\mathbf{x},t)\right)^T,$$

$$\mathbf{n}^{(eq)}(\mathbf{x},t)=\left(n_0^{(eq)}(\mathbf{x},t),n_1^{(eq)}(\mathbf{x},t),\cdots,n_4^{(eq)}(\mathbf{x},t)\right)^T,$$

where $g_i(\mathbf{x},t)$ is the temperature distribution function, $\mathbf{n}(\mathbf{x},t)$ and $\mathbf{n}^{(eq)}(\mathbf{x},t)$ are the velocity moments of the discrete distribution functions $\mathbf{g}$ and the corresponding equilibrium moments, respectively.

In the D2Q5 model, the five discrete velocities $\{\mathbf{e}_i \mid i = 0, 1, \cdots, 4\}$ are given by

$$\mathbf{e}_i = \begin{cases} (0,0), & i=0 \\ \left(\cos\left[(i-1)\pi/2\right], \sin\left[(i-1)\pi/2\right]\right)c, & i = 1 \sim 4 \end{cases}. \tag{B.3}$$

For the D2Q5 model, the transformation matrix $\mathbf{N}$ is given by

$$\mathbf{N} = \begin{pmatrix} 1 & 1 & 1 & 1 & 1 \\ 0 & 1 & 0 & -1 & 0 \\ 0 & 0 & 1 & 0 & -1 \\ -4 & 1 & 1 & 1 & 1 \\ 0 & 1 & -1 & 1 & -1 \end{pmatrix}. \tag{B.4}$$

The transformation matrix $\mathbf{N}$ linearly transforms the discrete distribution functions $\mathbf{g} \in \mathbb{V} = \mathbb{R}^5$ to their velocity moments $\mathbf{n} \in \mathbb{M} = \mathbb{R}^5$:

$$\mathbf{n} = \mathbf{N}\mathbf{g}, \quad \mathbf{g} = \mathbf{N}^{-1}\mathbf{n}. \tag{B.5}$$

In the system of $\{g_i\}$, only the temperature $T \equiv n_0 = \sum_{i=0}^{4} g_i$ is conserved quantity. The equilibrium moments $\{n_i^{(eq)} \mid i = 0, 1, \cdots, 4\}$ for the velocity moments $\{n \mid i = 0, 1, \cdots, 4\}$ are defined as [44]

$$n_0^{(eq)} = T, \quad n_1^{(eq)} = \frac{u_x T}{\sigma}, \quad n_2^{(eq)} = \frac{u_y T}{\sigma}, \quad n_3^{(eq)} = \varpi T, \quad n_4^{(eq)} = 0, \tag{B.6}$$

where $\varpi$ is a constant ($-4 < \varpi < 1$). In the present work, $\varpi$ is set to be $0$. The diagonal relaxation matrix $\Theta$ is given by:

$$\Theta = \text{diag}(1, \zeta_1, \zeta_2, \zeta_3, \zeta_4). \tag{B.7}$$

The effective thermal diffusivity $\alpha_e$ is defined as $\alpha_e = \sigma c_{sT}^2 (\tau_T - 0.5)\delta_t$ with $\zeta_1 = \zeta_2 = 1/\tau_T$ and $c_{sT}^2 = (4+\varpi)c^2/10$.

# Figure Captions

Fig. 1. Discrete velocities of the D2Q8 model.

Fig. 2. Velocity profiles of the porous Poiseuille flow for different Re and Da with $\phi = 0.1$ and $J = 1$. Solid lines represent the FD solutions [22] and symbols represent the present MRT-LB results.

Fig. 3. Velocity profiles of the porous Couette flow for different Re and Da with $\phi = 0.1$ and $J = 1$. Solid lines represent the FD solutions [22] and symbols represent the present MRT-LB results.

Fig. 4. Velocity profiles of the porous Couette flow for different viscosity ratio J with Re = 0.01 and Da = 0.001. Solid lines represent the analytical solutions from Eq. (25) and symbols represent the present MRT-LB results.

Fig. 5. Streamlines of the lid-driven flow in a porous cavity for different Da with $\phi = 0.1$ and $Re = 10$: (a) $Da = 10^{-2}$; (b) $Da = 10^{-3}$; (c) $Da = 10^{-4}$.

Fig. 6. Velocity profiles through the center of the cavity: (a) horizontal velocity component $u_x$ in the vertical midplane $x = L/2$; (b) vertical velocity component $u_y$ in the horizontal midplane $y = L/2$. Solid lines represent the FD solutions [22] and symbols represent the present MRT-LB results.

Fig. 7. Velocity profiles through the center of the cavity: (a) horizontal velocity component $u_x$ in the vertical midplane $x = L/2$; (b) vertical velocity component $u_y$ in the horizontal midplane $y = L/2$. Solid lines represent the benchmark solutions [42] and symbols represent the present MRT-LB results.

Fig. 8. Natural convection in a square cavity filled with a porous medium.

Fig. 9. Streamlines and isotherms for different Ra and Da with $Pr = 1$ and $\phi = 0.6$: (a) $Da = 10^{-2}$, $Ra = 10^5$; (b) $Da = 10^{-4}$, $Ra = 10^7$; (c) $Da = 10^{-6}$, $Ra = 10^9$.

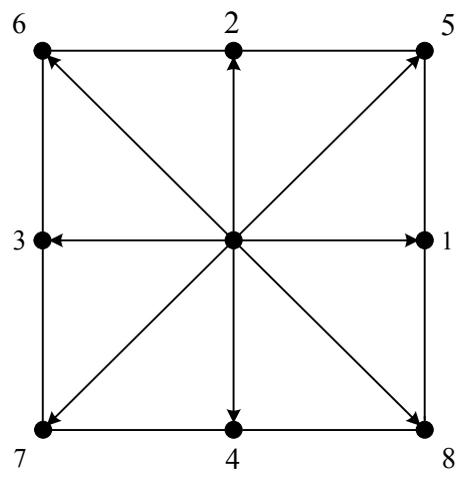

Fig. 1. Discrete velocities of the D2Q8 model.

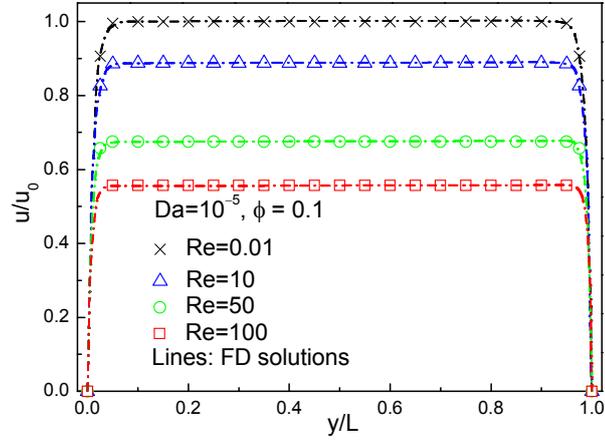

(a)

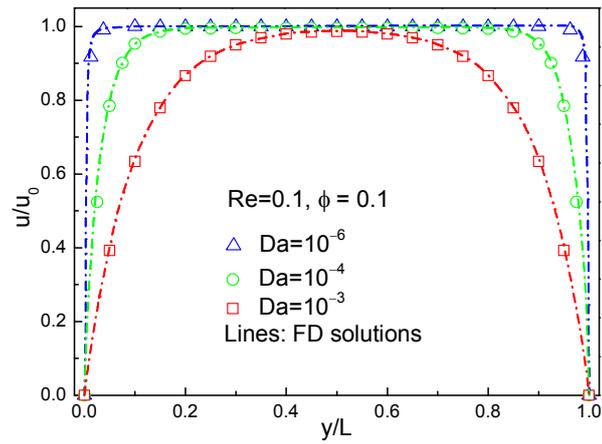

(b)

Fig. 2. Velocity profiles of the porous Poiseuille flow for different Re and Da with $\phi = 0.1$ and $J = 1$. Solid lines represent the FD solutions [22] and symbols represent the present MRT-LB results.

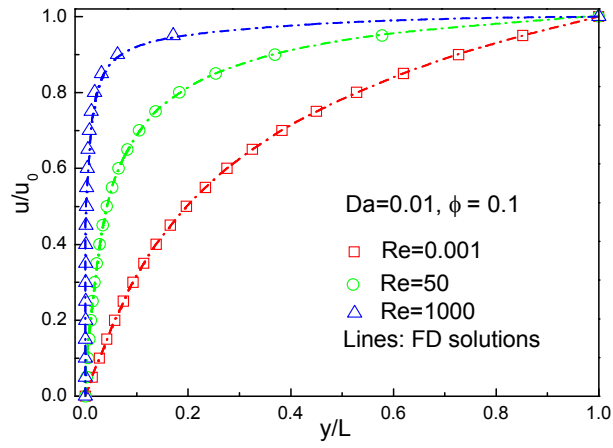

(a)

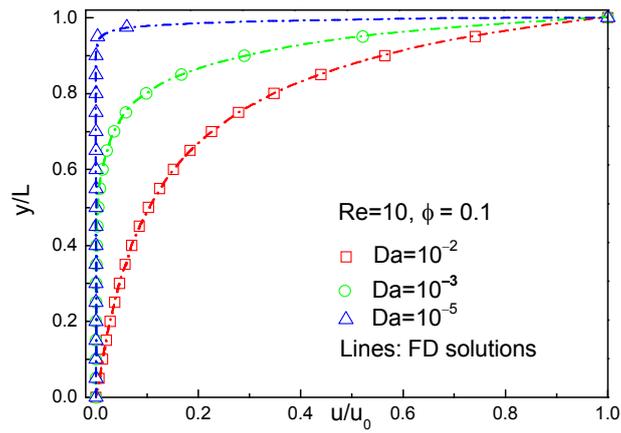

(b)

Fig. 3. Velocity profiles of the porous Couette flow for different Re and Da with $\phi = 0.1$ and $J = 1$. Solid lines represent the FD solutions [22] and symbols represent the present MRT-LB results.

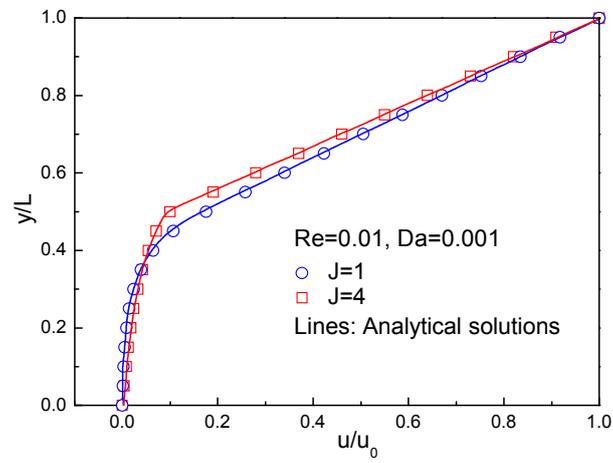

Fig. 4. Velocity profiles of the porous Couette flow for different viscosity ratio $J$ with $Re = 0.01$ and $Da = 0.001$. Solid lines represent the analytical solutions from Eq. (25) and symbols represent the present MRT-LB results.

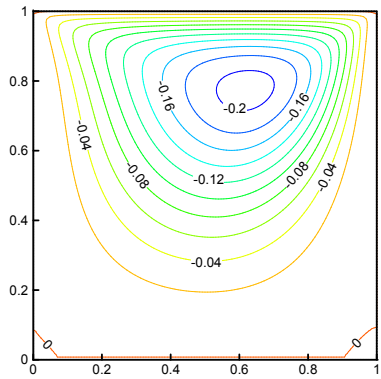

(a) $Da = 10^{-2}$

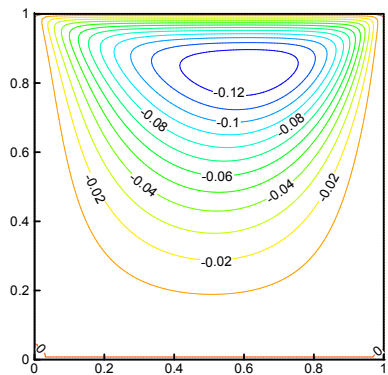

(b) $Da = 10^{-3}$

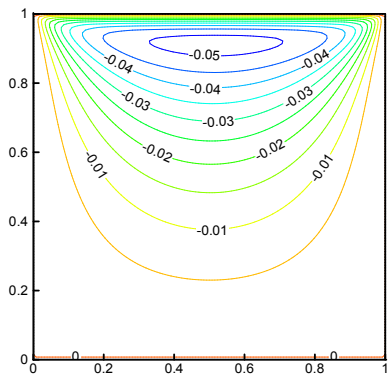

(c) $Da = 10^{-4}$

Fig. 5. Streamlines of the lid-driven flow in a porous cavity for different $Da$ with $\phi = 0.1$ and $Re = 10$: (a) $Da = 10^{-2}$; (b) $Da = 10^{-3}$; (c) $Da = 10^{-4}$.

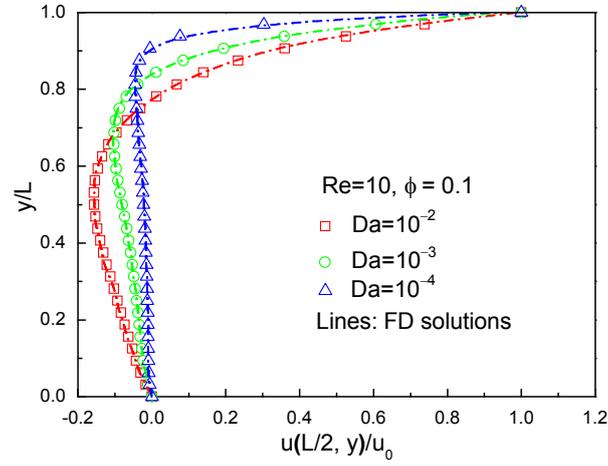

(a)

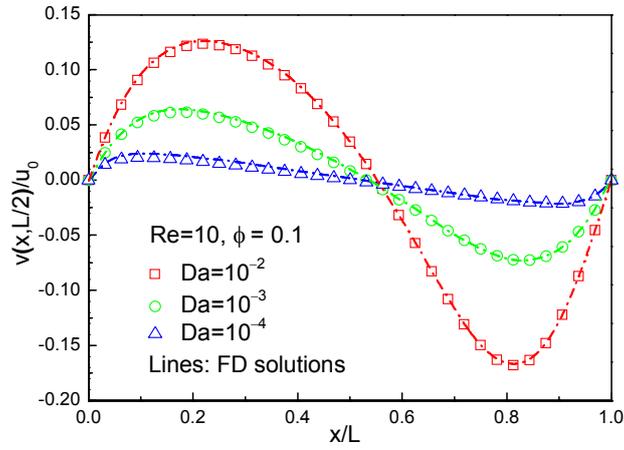

(b)

Fig. 6. Velocity profiles through the center of the cavity: (a) horizontal velocity component $u_x$ in the vertical midplane $x = L/2$; (b) vertical velocity component $u_y$ in the horizontal midplane $y = L/2$. Solid lines represent the FD solutions [22] and symbols represent the present MRT-LB results.

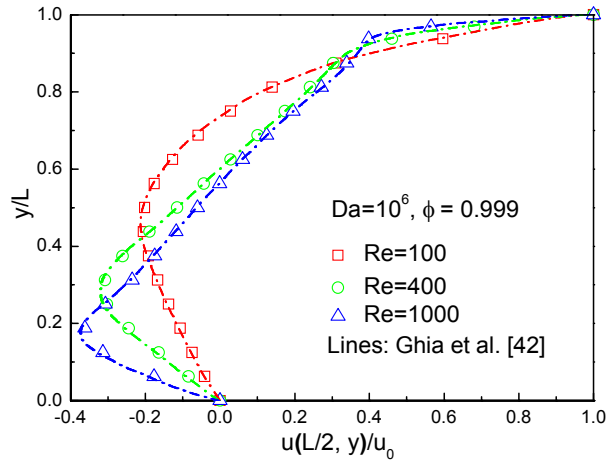

(a)

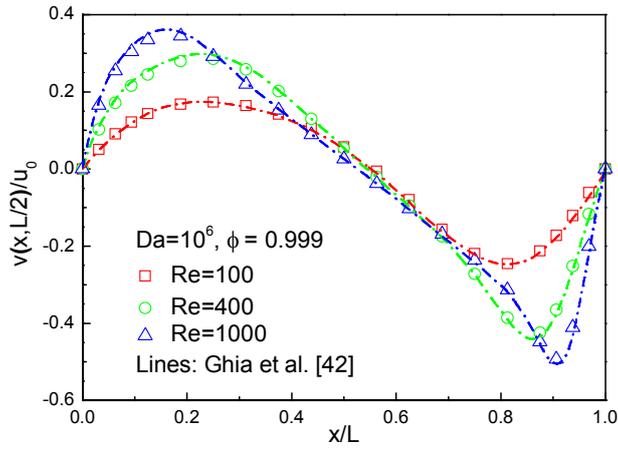

(b)

Fig. 7. Velocity profiles through the center of the cavity: (a) horizontal velocity component $u_x$ in the vertical midplane $x = L/2$; (b) vertical velocity component $u_y$ in the horizontal midplane $y = L/2$. Solid lines represent the benchmark solutions [42] and symbols represent the present MRT-LB results.

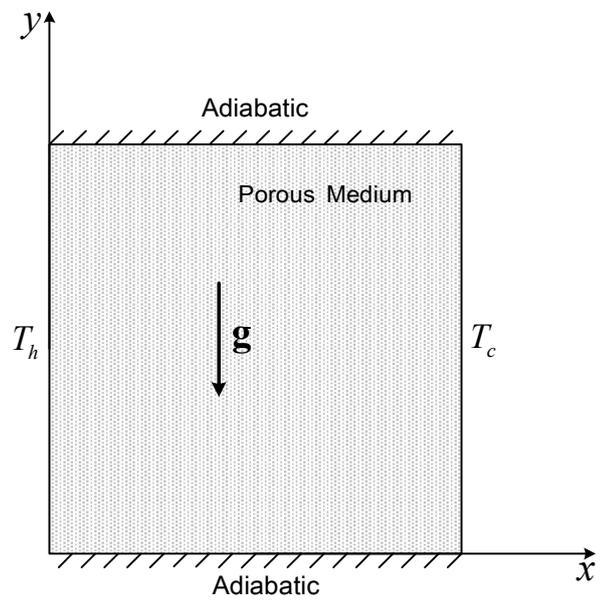

Fig. 8. Natural convection in a square cavity filled with a porous medium.

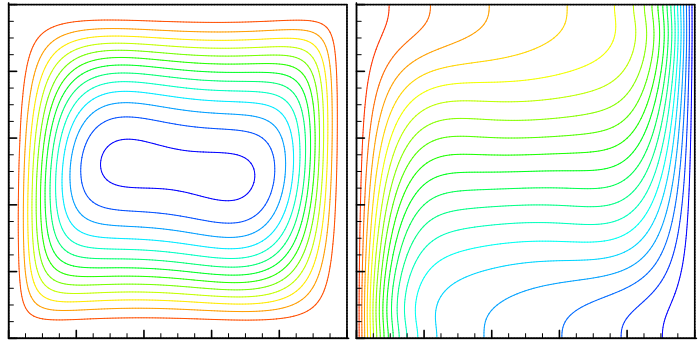

(a) $Da = 10^{-2}$, $Ra = 10^{5}$

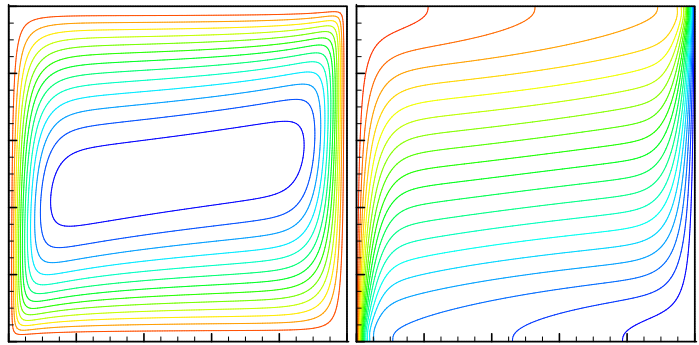

(b) $Da = 10^{-4}$, $Ra = 10^{7}$

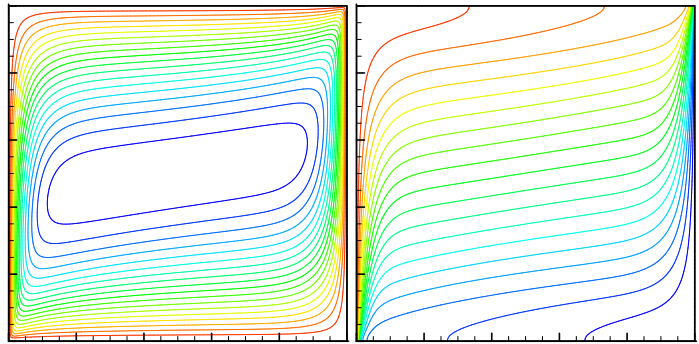

(c) $Da = 10^{-6}$, $Ra = 10^{9}$

Fig. 9. Streamlines and isotherms for different $Ra$ and $Da$ with $Pr = 1$ and $\phi = 0.6$ : (a) $Da = 10^{-2}$, $Ra = 10^{5}$; (b) $Da = 10^{-4}$, $Ra = 10^{7}$; (c) $Da = 10^{-6}$, $Ra = 10^{9}$.

**Table Caption**

Table 1. Comparisons of the average Nusselt numbers for different $Ra$, $Da$ and $\phi$ with $Pr = 1.0$.

| Da | Ra | $\phi = 0.4$ | | | $\phi = 0.6$ | | | $\phi = 0.9$ | | |
|---|---|---|---|---|---|---|---|---|---|---|
| | | Ref.[4] | Ref.[7] | Present | Ref.[4] | Ref.[7] | Present | Ref.[4] | Ref.[7] | Present |
| $10^{-2}$ | $10^3$ | 1.010 | 1.008 | 1.008 | 1.015 | 1.012 | 1.012 | 1.023 | - | 1.018 |
| | $10^4$ | 1.408 | 1.359 | 1.364 | 1.530 | 1.489 | 1.495 | 1.640 | - | 1.637 |
| | $10^5$ | 2.983 | 2.986 | 3.005 | 3.555 | 3.430 | 3.451 | 3.910 | - | 3.930 |
| $10^{-4}$ | $10^5$ | 1.067 | 1.064 | 1.065 | 1.071 | 1.066 | 1.068 | 1.072 | - | 1.070 |
| | $10^6$ | 2.550 | 2.580 | 2.605 | 2.725 | 2.686 | 2.714 | 2.740 | - | 2.798 |
| | $10^7$ | 7.810 | 7.677 | 7.770 | 8.183 | 8.452 | 8.540 | 9.202 | - | 9.244 |
| $10^{-6}$ | $10^7$ | 1.079 | 1.074 | 1.076 | 1.079 | 1.074 | 1.076 | 1.08 | - | 1.077 |
| | $10^8$ | 2.970 | 2.969 | 3.020 | 2.997 | 2.982 | 3.037 | 3.00 | - | 3.041 |
| | $10^9$ | 11.460 | 11.699 | 11.476 | 11.790 | 12.098 | 11.878 | 12.01 | - | 12.142 |